\documentclass[12pt]{article}
\usepackage{authblk}
\usepackage{fontenc}
\usepackage[utf8]{inputenc}
\usepackage{amsmath,amssymb}
\usepackage{graphicx}

\title{\small\textbf{lNHOMOGENEOUS VACUUM STATES IN NAMBU-JONA-LASINIO MODEL}}
\author{Sergii Kutnii}
\affil{\small\textit{Bogolyubov Institute for Theoretical Physics, Kyiv, Ukraine}}

\begin{document}

\maketitle

\begin{abstract}
The mean field approach to the Nambu-Jona-Lasinio model is developed systematically. Approximate mean field action is obtained, based on the study of divergencies in the mean field action. A special scalar case of the approximate motion equations is studied and inhomogenouous solutions are discussed. It is shown that the model can have inhomogeneous vacuum configurations which leads to bound fermionic states.
\end{abstract}

\section{\small INTRODUCTION}

The Nambu-Jona-Lasinio model\cite{NJL} has been proposed in 1961, long before QCD arose. 
It is a common belief that the model can be derived from QCD in a low-energy limit. An example of such derivation is presented in \cite{Volkov}. However, it relies on unproven (though plausible) assumptions about the structure of gluon propagator in the low-energy limit.

In a previous paper\cite{Me} it has been shown that NJL can be derived from QCD using the mean field approach developed by Kondo \cite{Kondo1, Kondo2}. It's notable that in this approach NJL is obtained naturally together with a cutoff parameter $\Lambda$ which is just the effective gluon mass generated by a dynamic mechanism suggested by Kondo in the works mentioned.
This solves the regularization problem in a way that is in perfect agreement with the belief that non-renormalizable models like NJL can only be effective field theories.

It has been demonstrated by Nambu and Jona-Lasinio that a simple model with only scalar and pseudoscalar quartic terms in its lagrangian has nontrivial vacuum state with fermionic condensate. Thus, initially massless fermions gain mass dynamically. This result was obtained in the mean field approach assuming spacial homogeneity, i.e. that the mean field has a constant value over the spacetime. However it has been pointed out recently that there can exist inhomogeneous vacuum configurations in this model. basar, Dunne and Thies have done this for $NJL$ model in 1+1 dimensions \cite{Thies}.

However, if it comes to vacuum inhomogeneities, mean field approach becomes tricky since it's necessary to solve the non-simplified gap equation. Therefore, it becomes interesting to develop some approximations that would be less tough but allowing to grasp vacuum inhomogeneities. This is done in this paper.

In the first part of this work NJL derivation from QCD is sketched briefly. Then the mean field approach to the model obtained is developed systematically. An approximate mean field action is obtained, based on the study of divergencies in the mean field action. Afterwards a special scalar case of the approximate motion equations is studied and inhomogenouous solutions are discussed.

\section{\small  NJL MODEL IN QUANTUM CHROMODYNAMICS}

We start with a plain (i.e.) flavorless SU(N) model

\begin{eqnarray}
 \mathcal{L} &=& \bar{\psi}\left(i\widehat{D} - m\right)\psi - \frac{1}{2}TrF_{\mu\nu}F^{\mu\nu} \nonumber\\
 D_\mu &=& \partial_\mu - igA_\mu \nonumber\\
 F_{\mu\nu} &=& \partial_\mu{}A_\nu - \partial_\nu{}A_\mu - ig\left[A_\mu,A_\nu\right]_{-}
\end{eqnarray}

 The cubic gluon self-interaction term can be removed by introducing an antisymmetric auxiliary field $B^a_{\mu\nu}$ in such a way that leads to the following lagrangian:

\begin{eqnarray}
 &&\mathcal{L} = Tr\left[-\frac{1 + \sigma^{-2}}{2}\left(
\partial_{\mu}A_{\nu} - \partial_{\mu}A_{\nu}\right)^2 
+ \frac{\left(1 + \sigma^2\right)g^2}{2}\left[A_{\mu},A_{\nu}\right]^2 -\right.\nonumber\\ &&
\left. -\sigma^{-1}B^{\mu\nu}\left(\partial_{\mu}A_{\nu} - \partial_{\mu}A_{\nu}\right) - 
i\sigma{}gB^{\mu\nu}\left[A_{\mu},A_{\nu}\right] - \frac{1}{2}B^2\right],
\end{eqnarray}
where $\sigma$ is an arbitrary numeric parameter and fermionic terms have been omitted for a moment. 

It's easy to demonstrate that

\begin{eqnarray}
& &\int\mathcal{D}B\exp\left\{-i\int{}d^4xTr\left[-\frac{1 + \sigma^{-2}}{2}\left(
\partial_{\mu}A_{\nu} - \partial_{\mu}A_{\nu}\right)^2 +\right.\right.\nonumber\\
&+&\left.\left. \frac{\left(1 + \sigma^2\right)g^2}{2}\left[A_{\mu},A_{\nu}\right]^2 - 
\sigma^{-1}B^{\mu\nu}\left(\partial_{\mu}A_{\nu} - \partial_{\mu}A_{\nu}\right) - 
i\sigma{}gB^{\mu\nu}\left[A_{\mu},A_{\nu}\right] - \frac{1}{2}B^2\right]\right\} = \nonumber\\
&=&\exp\left\{-i\int{}d^4x\left[-\frac{1}{2}TrF^2\right]\right\}\int\mathcal{D}B\exp\left\{-i\int{}d^4x\left[-\frac{1}{2}TrF^2 -\right.\right.\nonumber\\
&-&\left.\left. \frac{1}{2}Tr\left\{
B_{\mu\nu} + \sigma^{-1}\left(\partial_{\mu}A_{\nu} - \partial_{\mu}A_{\nu}\right) + 
i\sigma{}g\left[A_\mu,A_\nu\right]\right\}^2\right]\right\} = \nonumber\\
& = &C\exp\left\{-i\int{}d^4x\left[-\frac{1}{2}TrF^2\right]\right\}
\end{eqnarray}

In order to get rid of of the fourth order gluon term we note that 

\begin{eqnarray}
Tr\left[A_\mu,A_\nu\right]^2 &=& TrA_\mu\left[A_\nu,\left[A^\mu,A^\nu\right]\right] = 
TrA_\mu\left(\left\{A_\nu,\left\{A^\mu,A^\nu\right\}\right\} - \left\{A^\mu,\left\{A_\nu,A^\nu\right\}\right\}\right) = \nonumber\\
&=& Tr\left\{A_\mu,A_\nu\right\}^2 - Tr\left\{A_\mu,A^\mu\right\}^2 = \nonumber\\
&=& Tr\left[\left\{A_\mu,A_\nu\right\} - \frac{\eta_{\mu\nu}}{4}\left\{A_\lambda,A^\lambda\right\}\right]^2 - 
\frac{3}{4}Tr\left\{A_\mu,A^\mu\right\}^2
\end{eqnarray}

Then we introduce auxiliary fields $\Theta_{\mu\nu} = \theta_{\mu\nu} + \tau^a_{\mu\nu}T^a 
(\Theta^\lambda_{\hspace{8pt}\lambda} = 0)$ and
$\Phi = \phi + \varphi^aT^a$ in order to arrive at 

\begin{eqnarray}
 &&\mathcal{L} = Tr\left[-\frac{1 + \sigma^{-2}}{2}\left(
\partial_{\mu}A_{\nu} - \partial_{\mu}A_{\nu}\right)^2 - \Theta^{\mu\nu}\left(\left\{A_\mu,A_\nu\right\} - \frac{\eta_{\mu\nu}}{4}
\left\{A_\lambda,A^\lambda\right\}\right) -\right.\nonumber\\ 
 &&- \Phi\left\{A_\mu,A^\mu\right\} - \left. \sigma^{-1}B^{\mu\nu}\left(\partial_{\mu}A_{\nu} - \partial_{\mu}A_{\nu}\right) - 
i\sigma{}gB^{\mu\nu}\left[A_{\mu},A_{\nu}\right] -\right.\nonumber\\ 
 &&-\left. \frac{1}{2}B^2 - 
\frac{1}{2g^2\left(1 + \sigma^2\right)}\Theta^2 + \frac{2}{3g^2\left(1 + \sigma^2\right)}\Phi^2\right]
\end{eqnarray}

It's obvious now that the Faddeev-Popov lagrangian for gluon fields can be transformed into

\begin{equation}
\mathcal{L} = A^{\mu\,a}\mathcal{K}_{\mu\nu}^{ab}A^{\nu\,b} + \mathcal{J}^a_\mu(B)A^{a\,\mu} + \mathcal{L}_{FP} +
\mathcal{L}_{MF}(\Phi,\Theta,B) + \bar{\psi}\left(i\widehat{D} - m_0\right) \psi \label{eq:asquare}                    
\end{equation}
, where
$\mathcal{L}_{FP}$ is the ghost part of the lagrangian and $\mathcal{L}_{MF}(\Phi,\Theta,B)$ 
depends on auxiliary fields only. The $\mathcal{K}$ operator is

\begin{eqnarray}
 \mathcal{K}_{\mu\nu}^{ab} &=& \frac{1 + \sigma^2}{2}\eta_{\mu\nu}\delta^{ab}\partial^2 - 
\frac{\delta^{ab}}{2}\left(1 + \sigma^2 + \xi^{-1}\right)\partial_\mu\partial_\nu - 
\frac{d^{abc}}{2}\tau^c_{\mu\nu} - \frac{\delta^{ab}}{2}\theta_{\mu\nu} - \nonumber\\ &-&
\frac{\eta_{\mu\nu}}{2}\left(d^{abc}\varphi^c + \delta^{ab}\phi\right) + 
\frac{\sigma{}g}{2}f^{abc}B^c_{\mu\nu},
\end{eqnarray} 
where $\xi$ is a gauge-fixing parameter.

The lagrangian now is quadratic in gluon fields. Thus the gluons can be integrated out and the effective action becomes

\begin{eqnarray}
 &&S_{eff} = -iTr\ln{i\mathcal{K}} + \nonumber\\ &&+ \int{}d^4xTr\left[- \frac{1}{2}B^2 - 
\frac{1}{2g^2\left(1 + \sigma^2\right)}\Theta^2 + \frac{2}{3g^2\left(1 + \sigma^2\right)}\Phi^2\right] 
-\nonumber\\&&- \mathcal{J}^a_\mu\left(\mathcal{K}^{-1}\right)^{ab,\mu\nu}\mathcal{J}^b_{\nu}.
\end{eqnarray}

The current is

\begin{equation}
 \mathcal{J}^a_\mu = J^a_\mu\left(B, C\right) + g\bar{\psi}\gamma_\mu{}T^a\psi,
\end{equation}
where $J^a_\mu\left(B, C\right)$ is made of the $B$ field and Faddeev-Popov ghosts .

It's easy to find the gluon condensate now.  Let us put $B = 0,\Theta = 0, \varphi^a = 0,
\phi = \phi_0 \neq 0$ which is the case with highest unbroken symmetry. Let us also fix the gauge as follows:
\begin{equation}1 + \sigma^2 + \xi^{-1} = 0.\label{eq:defgauge}\end{equation}

Then the equation for $\phi_0$ follows immediately:

\begin{equation}\frac{\phi_0}{3g^2\left(1 + \sigma^2\right)} - 
i\int\frac{d^4p}{\left(1 + \sigma^2\right)p^2 + \phi_0} = 0\end{equation}
 
Putting 
\begin{equation}\sigma = i\rho, \rho > 1\end{equation}
we obtain a nontrivial solution

\begin{equation}
\phi_0 = \pm\frac{\rho^2 -1}{\pi{}g\sqrt{6I}}.
\end{equation}

Wick rotation was made here so 

\begin{equation}I = \int\limits_0^\infty\frac{z^3dz}{z^2 + 1}\end{equation}
 This integral is divergent and needs regularization but we won't go into it.

Having this solution, we can build the mean field approximation which is equivalent to leaving the zero order term only in the saddle-point expansion of the path integral that corresponds to lagrangian (\ref{eq:asquare}). The mean field lagrangian is

\begin{eqnarray}
 \mathcal{L} &=& -TrA^\mu\left[\eta_{\mu\nu}\left(\rho^2 - 1\right)\partial^2
 + \left(1 - \rho^2 +\xi^{-1}\right)\partial_\mu\partial_\nu+ \eta_{\mu\nu}\phi_0\right]A^\nu 
 + \nonumber\\
& + &\bar{\psi}\left(i\widehat{D} - m_0\right)\psi + \mathcal{L}_{FP} \label{eq:gluon_meanfield}
\end{eqnarray}

It's clear now that our solution describes dynamic gluon mass generation:

\begin{equation}
 m_A^2 = \frac{\phi_0}{\rho^2 - 1}
\end{equation}
Note that the effective mass doesn't depend on the free parameter $\rho$.

Let us fix the gauge (\ref{eq:defgauge}) in the action (\ref{eq:gluon_meanfield}) and integrate out the gluons omitting the Faddev-Popov ghosts. The resulting action will be 

\begin{eqnarray}
 &&S = \int{}d^4x\bar{\psi}\left(i\widehat{\partial} - m_0\right)\psi + \nonumber\\
&&+ \frac{1}{2\left(\rho^2 - 1\right)}\int{}d^4xd^4y\bar{\psi}(x)\gamma_\mu{}T^a\psi(x)\langle{x}|\left(\partial^2 + m_A^2\right)^{-1}|y\rangle\bar{\psi}(y)\gamma^\mu{}T^a\psi(y)\nonumber\\
\end{eqnarray}

The Green's function in the second term of the action can be expanded as follows:

\begin{equation}
 \langle{x}|\left(\partial^2 + m_A^2\right)^{-1}|y\rangle = 
\frac{\delta(x - y)}{m_A^2}\sum\limits_{n = 0}^{\infty}\frac{(-1)^n}{m_A^{2n}}\partial^{2n}
\end{equation}

Therefore, we can conclude that the higher order terms will be small and the expansion can converge if 
\begin{equation}|p_{\mu}| < m_A \equiv \Lambda \label{eq:defapprox}\end{equation} 
This condition defines the low energy limit of the theory. 

The zero order term of the expansion above gives a four-fermion term:
\begin{equation}\mathcal{L}_{qq}^{(0)} = \frac{1}{2\left(\rho^2 - 1\right)\Lambda^2}\bar{\psi}_{i}\gamma_{\mu}T^{a}_{ij}\psi_{j}\bar{\psi}_{k}\gamma^{\mu}T^{a}_{kl}\psi_{l}\end{equation}

Using the technique developed in \cite{Me}, this term can be simplified further which leads to the following fermionic lagrangian: 

\begin{eqnarray}
&&\mathcal{L}_q = \bar{\psi}_{i}\left(i\hat{\partial} - m_0\right)\psi_i
-\nonumber\\&& + \frac{1}{4\left(\rho^2 - 1\right)\Lambda^2}
\left[\bar{\psi}_i\psi_i\bar{\psi}_k\psi_k - \bar{\psi}_i\gamma^5\psi_i\bar{\psi}_k\gamma^5\psi_k\right. 
\nonumber\\ &&- \left. 
\frac{N + 2}{2N}\bar{\psi}_{i}\gamma_{\mu}\psi_{i}\bar{\psi}_{k}\gamma^{\mu}\psi_{k} - 
\frac{1}{2}\bar{\psi}_{i}\gamma^5\gamma_{\mu}\psi_{i}\bar{\psi}_{k}\gamma^5\gamma^{\mu}\psi_{k}\right]
\label{eq:NJLagrangian}
\end{eqnarray}

Diagonality with respect to color indices is a notable feature of this lagrangian. 
It allows us to define colorless mean fields which will be done below.

\section{\small DYNAMICS OF THE FERMIONIC MEAN FIELDS}

Let us now study the NJL model itself. We'll work within the mean field approach to it developed first by Nambu and Jona-Lasinio to find the vacuum condensate of the model. Let us start with the following path integral that corresponds to the lagrangian (\ref{eq:NJLagrangian}):
\begin{eqnarray}
 Z &=& \int\mathcal{D}\bar{\psi}\mathcal{D}\psi\exp\left\{-i\int\,d^4x\left[\bar{\psi}\left(i\hat{\partial} - m_0\right)\psi + 
\alpha\left[\bar{\psi}\psi\bar{\psi}\psi - \bar{\psi}\gamma^5\psi\bar{\psi}\gamma^5\psi\right.\right.\right. \nonumber\\ 
&-& \left. \left . \left. \beta\bar{\psi}\gamma_{\mu}\psi\bar{\psi}\gamma^{\mu}\psi + 
\frac{1}{2}\bar{\psi}\gamma^5\gamma_{\mu}\psi\bar{\psi}\gamma^5\gamma^{\mu}\psi\right]\right]\right\},
\label{eq:continual}
\end{eqnarray}
\noindent Where we've put
\begin{eqnarray}
\alpha &=& \frac{1}{4\left(\rho^2 - 1\right)\Lambda^2}\\
\beta &=& \frac{N + 2}{2N}\label{eq:deflambda}
\end{eqnarray}
for the sake of brevity.

Now we introduce one scalar, one pseudoscalar, one vector and one pseudovector auxiliary fields  $\xi, \eta, v^{\mu}, w^{\mu}$ to get rid of the fourth-order terms. We arrive at the following path integral:
\begin{eqnarray}
 Z &=& \int\mathcal{D}\bar{\psi}\mathcal{D}\psi\mathcal{D}\xi\mathcal{D}\eta\mathcal{D}v\mathcal{D}w\exp\left\{-i\int\,d^4x\left[\bar{\psi}\left(i\hat{\partial} - m_0\right)\psi + 
\xi\bar{\psi}\psi + \eta\bar{\psi}\gamma^5\psi\right.\right. \nonumber\\ 
&+&  \left . \left. v_{\mu}\bar{\psi}\gamma^{\mu}\psi + w_\mu\bar{\psi}\gamma^5\gamma^{\mu}\psi - 
\frac{1}{4\alpha}\left(\xi^2-\eta^2 - \beta^{-1}v^2 + 2w^2\right)\right]\right\} = \nonumber\\
&=&\int\mathcal{D}\bar{\psi}\mathcal{D}\psi\mathcal{D}\xi\mathcal{D}\eta\mathcal{D}v\mathcal{D}w
\exp\left\{-i\int\,d^4x\left[\bar{\psi}\left(i\hat{\partial} - m_0 + \xi + \eta\gamma^5 + \widehat{v} + \gamma^5\widehat{w}\right)\psi -\right.\right.\nonumber\\  
&-&\left. \left. \frac{1}{4\alpha}\left(\xi^2-\eta^2 - \beta^{-1}v^2 + 2w^2\right)\right]\right\}
\label{eq:NJLMF}
\end{eqnarray}

Let us now define the matrix mean field to be
\begin{equation}\widehat{\Omega} = \xi + \eta\gamma^5 + \widehat{v} + \gamma^5\widehat{w}.\label{eq:defomega}\end{equation}

It's obvious that the terms in (\ref{NJLMF}) that are  quadratic in auxiliary fields can be understood as 
some bilinear function of $\Omega$, therefore we'll denote them simply as $\Phi_2(\widehat{\Omega}).$

The fermions can be integrated out now and we get the following effective action for $\widehat{\Omega}$:

\begin{equation}S_{eff}\left[\Omega\right] = - iNTr\ln{i\left(i\hat{\partial} - m_0 + \widehat{\Omega}\right)} 
- \frac{1}{4\alpha}\int\,d^4x \Phi_2(\widehat{\Omega})\label{eq:effact}\end{equation} 
The equation of motion that can be derived from it is
\begin{equation}\frac{1}{4\alpha}\frac{\partial\Phi_2\left(\widehat{\Omega}\right)}{\partial\widehat{\Omega}}(x) = \widehat{G}_{\Omega}(x,x)\label{eq:Gorkov}
\end{equation}
where $\widehat{G}_{\Omega}(x,y)$ is the Green's function for operator $i\partial - m_0 + \widehat{\Omega}$. This equation can be called gap equation or Gor'kov equation, based on the analogy to superconductivity.
By putting $\widehat{\Omega} = const$ it can be reduced to a system of algebraic equations on the components of $\widehat{\Omega}$ which was in fact done by Nambu and Jona-Lasinio themselves in the very first article on the subject.

Studying the inhomogeneities, however, is much more complicated since one has to deal with a nonlinear integral equation (\ref{eq:Gorkov}). Another difficulty is that deriving the Green's function in the right-hand side of it encounters the problem of ultraviolet divergencies. Fortunately enough, we've got the cutoff, but it should be applied explicitly which is what we're going to do now.

Let us expand the logarithm in the effective action (\ref{eq:effact}). We've got
\begin{eqnarray}S_{eff}\left[\Omega\right] &=& -iNTr\ln{i\left(i\hat{\partial} - m_0\right)} - 
iNTr\ln\left(1 + \frac{1}{i\hat{\partial} - m_0}\circ\widehat{\Omega}\right) - 
\frac{1}{4\alpha}\int\,d^4x \Phi_2(\widehat{\Omega}) = \nonumber\\&=& 
- \frac{1}{4\alpha}\int\,d^4x \Phi_2(\widehat{\Omega}) - 
iNSp\widehat{S}(0)\int\,d^4x\widehat{\Omega}(x) +\nonumber\\ 
&+&\frac{iN}{2}Sp\int\,d^4xd^4y\widehat{S}(x-y)\widehat{\Omega}(y)\widehat{S}(y-x)\widehat{\Omega}(x)+...
\label{eq:series}
\end{eqnarray},
where $\widehat{S}(x-y)$ is the plain Dirac propagator. 

The first term in the expansion can be rewritten as follows:
\begin{equation}
Sp\widehat{S}(0)\int\,d^4x\widehat{\Omega}(x) = 
Sp\int\frac{d^4xd^4p}{(2\pi)^4}\frac{\hat{p}+m_0}{p^2-m_0^2}\widehat{\Omega}(x)
\end{equation}
Let us now demonstrate that each term in the series can be reduced to $$\int\,d^4pd^4x\sum\limits_{i = 0}^{\infty}\widehat{\Xi}_i(x,p)$$. 
For this, it's enough to demonstrate that
$$\int\,d^4y\widehat{S}(x-y)\widehat{\Omega}(y)\int\,d^4pe^{-ip(y-z)}\widehat{\Xi}(p,z) = 
\int\,d^4pe^{-ip(x-z)}\widehat{\widetilde{\Xi}}(p,z)$$
So
\begin{eqnarray}
&&\int\,d^4y\widehat{S}(x-y)\widehat{\Omega}(y)\int\,d^4pe^{-ip(y-z)}\widehat{\Xi}(p,z) =\nonumber\\&&= \int\frac{d^4yd^4qd^4p}{(2\pi)^4}e^{-iq(x-y)}\frac{\hat{q}+m_0}{{q}^2-m_0^2}\widehat{\Omega}(y)e^{-ip(y-z)}\widehat{\Xi}(p,z) = \nonumber\\
&&=\int\frac{d^4yd^4qd^4p}{(2\pi)^4}e^{-iq(x-y)}\frac{\hat{q}+m_0}{{q}^2-m_0^2}\times\nonumber\\&&\times
\sum\limits^{\infty}_{n=0}\frac{(y-z)^{\alpha_1}..(y-z)^{\alpha_n}}{n!}\left[\frac{\partial}{\partial{z}^{\alpha_1}}..\frac{\partial}{\partial{z}^{\alpha_n}}\widehat{\Omega}(z)\right]
e^{-ip(y-z)}\widehat{\Xi}(p,z) = \nonumber\\&&=
\int\frac{d^4yd^4qd^4p}{(2\pi)^4}e^{-iq(x-y)}\frac{\hat{q}+m_0}{{q}^2-m_0^2}\times\nonumber\\&&\times\sum\limits^{\infty}_{n=0}\frac{i^n}{n!}\left[\frac{\partial}{\partial{z}^{\alpha_1}}..\frac{\partial}{\partial{z}^{\alpha_n}}\widehat{\Omega}(z)\right]
\left[\frac{\partial}{\partial{p}_{\alpha_1}}..\frac{\partial}{\partial{p}_{\alpha_n}}e^{-ip(y-z)}\right]\widehat{\Xi}(p,z) = \nonumber\\&&=\int\frac{d^4yd^4qd^4p}{(2\pi)^4}e^{-iq(x-y)}\frac{\hat{q}+m_0}{{q}^2-m_0^2}e^{-ip(y-z)}\times\nonumber\\&&\times\sum\limits^{\infty}_{n=0}\frac{(-i)^n}{n!}\left[\frac{\partial}{\partial{z}^{\alpha_1}}..\frac{\partial}{\partial{z}^{\alpha_n}}\widehat{\Omega}(z)\right]
\frac{\partial}{\partial{p}_{\alpha_1}}..\frac{\partial}{\partial{p}_{\alpha_n}}\widehat{\Xi}(p,z)= \nonumber\\
&&=\int\,d^4qd^4pe^{-i(qx-pz)}\frac{\hat{q}+m_0}{{q}^2-m_0^2}\int\frac{d^4y}{(2\pi)^4}e^{-i(p-q)y}\times\nonumber\\&&\times\sum\limits^{\infty}_{n=0}\frac{(-i)^n}{n!}\left[\frac{\partial}{\partial{z}^{\alpha_1}}..\frac{\partial}{\partial{z}^{\alpha_n}}\widehat{\Omega}(z)\right]
\frac{\partial}{\partial{p}_{\alpha_1}}..\frac{\partial}{\partial{p}_{\alpha_n}}\widehat{\Xi}(p,z)=\nonumber\\&&=\int\,d^4qd^4pe^{-i(qx-pz)}\frac{\hat{q}+m_0}{{q}^2-m_0^2}\delta(p-q)\sum\limits^{\infty}_{n=0}\frac{(-i)^n}{n!}\left[\frac{\partial}{\partial{z}^{\alpha_1}}..\frac{\partial}{\partial{z}^{\alpha_n}}\widehat{\Omega}(z)\right]
\frac{\partial}{\partial{p}_{\alpha_1}}..\frac{\partial}{\partial{p}_{\alpha_n}}\widehat{\Xi}(p,z)=\nonumber\\&&=\int\,d^4pe^{-ip(x-z)}\frac{\hat{p}+m_0}{{p}^2-m_0^2}\sum\limits^{\infty}_{n=0}\frac{(-i)^n}{n!}\left[\frac{\partial}{\partial{z}^{\alpha_1}}..\frac{\partial}{\partial{z}^{\alpha_n}}\widehat{\Omega}(z)\right]\frac{\partial}{\partial{p}_{\alpha_1}}..\frac{\partial}{\partial{p}_{\alpha_n}}\widehat{\Xi}(p,z)\hspace{0.7cm}Q.E.D
\end{eqnarray}

Therefore, it's now obvious that if we define the operator 
\begin{equation}
 \widehat{V}(x,p) = \sum\limits^{\infty}_{n=0}\frac{(-i)^n}{n!}\left[\frac{\partial}{\partial{z}^{\alpha_1}}..\frac{\partial}{\partial{z}^{\alpha_n}}\widehat{\Omega}(z)\right]
\frac{\partial}{\partial{p}_{\alpha_1}}..\frac{\partial}{\partial{p}_{\alpha_n}}
\label{eq:vertex}
\end{equation}
it's possible to express the expansion (\ref{eq:series}) as
\begin{equation}
 S_{eff}\left[\Omega\right] = - \frac{1}{4\alpha}\int\,d^4x \Phi_2(\widehat{\Omega}) + 
iN\int\frac{d^4xd^4p}{(2\pi)^4}Sp\sum\limits_{k=1}^\infty\frac{(-1)^{k}}{k}\left[\frac{\widehat{p} + m_0}{p^2 - m_0^2}\widehat{V}(x,p)\right]^k\circ1
\label{eq:loop}
\end{equation}

We can easily read from this that all the integrations in (\ref{eq:series}) reduce to expressions of type
\begin{equation}
I(m,n)_{\alpha_1..\alpha_m} = 
\int\,d^4p\frac{p_{\alpha_1}..p_{\alpha_m}}{\left(p^2 - m_0^2\right)^n}
\label{eq:defint}
\end{equation}

First of all, it's easy to see that integrals like this are nonzero for even values of $m$ only and are finite if
\begin{equation}d(n,m) = 2n - m - 4 > 0\end{equation}. 
It's obvious that differentiation $\frac{\partial}{\partial{}p^{\mu}}$ of the integrand in (\ref{eq:defint}) increments its $d(n,m)$ by $1$ while leaving the difference $n-m$ unchanged. Thus we can conclude from (\ref{eq:loop}) that the series contains only the integrals (\ref{eq:defint}) with $m \leq n$. But it's also true that a $k$-th order in $\widehat{\Omega}$ term can contain only the integrals with $n \geq k$.

Therefore we immediately conclude that the expansion (\ref{eq:loop}) has only finite number of divergent terms which are
\begin{enumerate}
\item first order term;
\item second order up to second derivatives of $\widehat{\Omega}$;
\item third order up to first derivatives;
\item fourth order with no derivatives of $\widehat{\Omega}$.
\end{enumerate}

Regularization by cutoff will replace the divergencies with some finite factors of order $\Lambda^n$ in the cutoff parameter. Therefore, the finite part can be treated as a small correction to the divergent one. So by omitting the finite terms we can build an approximation of the effective action (\ref{eq:effact}) that has remarkable features. First, it's similar in structure to the Ginzburg-Landau functional in superconductivity, since it contains derivatives up to second and nonlinearities up to fourth order; second, it requires no extra conditions being imposed on the model to be valid.

To calculate the terms of our interest we use the Passarino-Veltman reduction \cite{PasVel}
\begin{eqnarray}
&&I(2m+1,n)_{\alpha_1..\alpha_{2m + 1}} = 0\nonumber\\
&&I(2m,n)_{\alpha_1..\alpha_{2m}} = CS\left(\eta_{\alpha_1\alpha_2}..\eta_{\alpha_{2m-1}\alpha_{2m}}\right)
\int{}d^4p\frac{p^{2m}}{\left[p^2 - m_0^2\right]^n}
\end{eqnarray}
where $C$ is a symmetry factor and $S\left(\eta_{\alpha_1\alpha_2}..\eta_{\alpha_{2m-1}\alpha_{2m}}\right)$ is a symmetric tensor power of the Lorentz metrics.

After rather long computation we obtain the following expression for the approximate "Ginzburg-Landau" action:
\begin{eqnarray}
\Xi &=& - \left(\rho^2 - 1\right)\Lambda^2\int{d^4x}\Phi_2(\Omega) - \nonumber\\
&&- \frac{N\ln{\left(\frac{\Lambda^2}{m^2} + 1\right)}}{32\pi^2}\int{}d^4xSp\left\{2m_0\left[Z(\Lambda) - m_0^2\right]\widehat{\Omega}\left(x\right) + m_0^2\left[\widehat{\Omega}(x)\right]^2 -\right.\nonumber\\
&&\left. - \frac{Z(\Lambda) - 2m_0^2}{4}\gamma_\mu\widehat{\Omega}(x)\gamma^\mu\widehat{\Omega}(x)
 + m_0\widehat{\Omega}(x)i\widehat{\partial}\widehat{\Omega}(x) - \right. \nonumber\\
&&\left.- \frac{1}{6}\left[\widehat{\partial}\widehat{\Omega}(x)\right]^2 - \frac{1}{12}\gamma_\mu\left(\partial_\nu\widehat{\Omega}(x)\right)\gamma^\mu\partial^\nu\widehat{\Omega}(x) - \right.\nonumber\\
&&\left.- \frac{m_0}{2}\widehat{\Omega}(x)\gamma_\mu\widehat{\Omega}(x)\gamma^\mu\widehat{\Omega}(x) + \right.\nonumber\\
&&\left. + \frac{i}{6}\gamma_\mu\widehat{\Omega}(x)\gamma^\mu\left[\partial_\nu\widehat{\Omega}(x)\right]\gamma^\nu\widehat{\Omega}(x) -
 \frac{i}{6}\gamma_\mu\widehat{\Omega}(x)\gamma^\mu\widehat{\Omega}(x)\widehat{\partial}\widehat{\Omega}(x) + \right.\nonumber\\
&&\left. + \frac{1}{24}Sp\left[\gamma_\mu\widehat{\Omega}(x)\gamma^\mu\widehat{\Omega}(x)\right]^2 
 + \frac{1}{48}Sp\gamma_\mu\widehat{\Omega}(x)\gamma_\nu\widehat{\Omega}(x)\gamma^\mu\widehat{\Omega}(x)\gamma^\nu\widehat{\Omega}(x)\right\},
\label{eq:action}
\end{eqnarray}
where it has been put
\begin{equation}
 Z(\Lambda) = \frac{\Lambda^2}{\ln\left(\frac{\Lambda^2}{m_0^2} + 1\right)}
\end{equation}
Note that differential and nonlinear terms are of the same logarithmic order in $\Lambda$. This implies that nonlinearities play an important role in the inhomogeneous case and can't be treated as just small corrections.

The only special case that should be discussed here is the zero initial mass case $m_0 = 0$ when infrared divergencies appear. This seems an obstacle but let us recall that this is in fact the case studied by Nambu and Jona-Lasinio where the vacuum condensate appears. Thus we can conclude that the infrared divergencies are an artifact of expansion around the false vacuum $\widehat{\Omega} = 0$ and the disease can be cured by expanding around the true vacuum of the theory. Putting $\widehat{\Omega} = \xi = const$ in the gap equation (\ref{eq:Gorkov}) we obtain
\begin{equation}
 \frac{N}{16\pi^2\left(\rho^2 - 1\right)}\left[1 - 
\xi^2\frac{\ln\left(\frac{\Lambda^2}{\xi^2} + 1\right)}{\Lambda^2}\right] = 1
\end{equation}
From this equation we can conclude that $\xi = \tau\Lambda$ where $\tau$ is a constant multiplier.
Let us recall the definition of $I(n,m)$ (\ref{eq:defint}). If we expand around the true vacuum $\widehat{\Omega} = \pm{M} = \pm\tau\Lambda$ then $I(n,m)$ get replaced with $\tilde(I)(n,m) = M^{-2n + m + 4}S(n,m)$ where $S(n,m)$ doesn't depend on $M$ and $\Lambda$. This can make our approximation even more plausible for the zero mass case than for nonzero mass since the finite terms get multiplied by $\Lambda^{-2n + m + 4}$.
The complete approximate action for the zero-mass case can be written out by making the following substitutions in (\ref{eq:action}):
$$\widehat{\Omega}\rightarrow\widehat{\widetilde{\Omega}};\\
 \Phi_2\left(\widehat{\Omega}\right) \rightarrow \Phi_2\left(\widehat{\widetilde{\Omega}} - M\right);\\
 Z(\Lambda) \rightarrow Z(\Lambda, M) = \frac{\Lambda^2}{\ln\left(\frac{\Lambda^2}{M^2} + 1\right)}
$$
\section{\small INHOMOGENEOUS SCALAR VACUUM}
It seems hard to study the action (\ref{eq:action}). However it posesses a simple scalar sector. 
By putting 
\begin{equation}
 \widehat{\Omega}(x) = \xi(x)\label{eq:defscalar}
\end{equation}
in the corresponding equation of motion the following equation can be obtained:
\begin{eqnarray}
 &&-G(N,\rho)Z(\Lambda)\xi(x) - 2m_0\left[Z(\Lambda) - m_0^2\right]   
+ 2\left[Z(\Lambda)- 3m_0^2\right]\xi(x) - \nonumber\\
&& - \Box\xi(x) + 6m_0\xi^2(x) - 2\xi^3(x) = 0,\label{eq:elliptic}
\end{eqnarray}
where $$G(N,\rho) = \frac{32\pi^2\left(\rho^2 - 1\right)}{N}$$.

The most interesting case here is $m_0 = 0$ when it reduces to the well-known $\phi^4$ equation:
\begin{equation}
\Box\xi(x) - \left[2 - G(N,\rho)\right]Z(\Lambda, M)\xi(x) + 2\xi^3(x) = 0 
\label{eq:phi4}
\end{equation}
This equation can be obtained by making the expansion around the true vacuum value of $\widehat{\Omega}$ discussed in the previous section and then changing the dependent variable back to $$\widehat{\Omega} = - M + \widehat{\widetilde{\Omega}}$$.

It has the famous one-dimensional kink solution

\begin{equation}
 \xi(z) = A\tanh\left(\kappa{z}\right),\label{eq:kink}
\end{equation}
which corresponds to a domain wall between the areas with different vacua.

The equation (\ref{eq:elliptic}) doesn't have the kink but it has a periodic solution that can be expressed in terms of elliptic functions which is similar to what was obtained by Thies et. al. It can also have a soliton solution as a special case.

\section{\small CONCLUSIONS}

We've studied the effective NJL mean field action. We've built the approximate action (\ref{eq:action}) that doesn't require any extra conditions to be imposed on the dynamic variables and is governed by the cutoff parameter only. So we can conclude that our approximation is valid under the same conditions as the NJL model itself.

The mean field dynamic in our approximation has a nontrivial scalar sector that is described with a variant of the $\phi^4$ equation in the zero initial mass case. Thus in this case we've got some nontrivial inhomogeneous vacuum configurations, in particular the kink solution.

Let us recall that the mean field approach allows us to construct an approximate fermionic lagrangian which in our case is 
\begin{equation}
\mathcal{L} = \bar{\psi}\left(i\widehat{\partial} - m_0 + \widehat{\Omega}\right)\psi
\end{equation} 
where we should substitute an actual solution of the gap equation for $\widehat{\Omega}$. Therefore we can see that vacuum inhomogeneities can lead to interesting fenomena like fermionic bound states. Indeed, if we put $m_0 = 0$ and $\widehat{\Omega} = \mu\tanh(z)$ which is just the kink case we get exactly the Jackiw-Rebbi problem\cite{JackiwRebbi} which is known to have the zero energy solution. We've studied the problem and found that it actually posesses discrete spectrum $E_n = \sqrt{n\left(2\mu - n\right)}$ but the details will be given in a subsequent paper.

The scalar sector studied here is only a small part of the picture. The approximation (\ref{eq:action}) can lead to other interesting fenomena.

The most thrilling question that arises here is whether there can exist stable compact 3D inhomogeneous vacuum configurations with bound fermions that can contribute to quark confinement.

\section{\small ACKNOWLEDGEMENTS}

We'd like to thank P. Holod for gudance D. Anchishkin for valuable comments and E. Gorbar for some article references.

\end{document}